\documentclass{aastex62}

\usepackage[utf8]{inputenc}

\received{February 3, 2019}
\revised{March 15, 2019}
\accepted{March 15, 2019}
\submitjournal{ApJ}

\shorttitle{Aurora during Maunder minimum}
\shortauthors{Isobe et al.}
\begin{document}

\title{Intense geomagnetic storm during Maunder minimum possibly by a quiescent filament eruption}

\correspondingauthor{Hiroaki Isobe}
\email{hi-isobe@kcua.ac.jp}

\author[0000-0002-0786-7307]{Hiroaki Isobe}
\affil{Faculty of Fine Arts, Kyoto City University of Arts \\
Kyoto 610-1197, Japan}
\affil{Unit of Synergetic Studies for Space, Kyoto University \\
Kyoto 606-8502, Japan}
\nocollaboration

\author{Yusuke Ebihara}
\affiliation{Research Institute for Sustainable Humanosphere, Kyoto University, Uji, 611-0011, Japan}
\affil{Unit of Synergetic Studies for Space, Kyoto University \\
Kyoto 606-8502, Japan}
\nocollaboration

\author{Akito D. Kawamura}
\affiliation{Astronomical Observatory, Graduate School of Science, Kyoto University, Kyoto 606-8502, Japan}

\author{Harufumi Tamazawa}
\affil{Faculty of Fine Arts, Kyoto City University of Arts \\
Kyoto 610-1197, Japan}
\affiliation{Disaster Prevention Research Institute, Kyoto University, Kyoto 611-0011}
\nocollaboration

\author{Hisashi Hayakawa}
\affiliation{Graduate School of Letters, Osaka University, Toyonaka, 5600043, Japan (JSPS Research Fellow)}
\affiliation{Rutherford Appleton Laboratory, Chilton, Didcot, Oxon OX11 0QX, UK}
\nocollaboration

%% Note that the \and command from previous versions of AASTeX is now
%% depreciated in this version as it is no longer necessary. AASTeX 
%% automatically takes care of all commas and "and"s between authors names.

%% AASTeX 6.2 has the new \collaboration and \nocollaboration commands to
%% provide the collaboration status of a group of authors. These commands 
%% can be used either before or after the list of corresponding authors. The
%% argument for \collaboration is the collaboration identifier. Authors are
%% encouraged to surround collaboration identifiers with ()s. The 
%% \nocollaboration command takes no argument and exists to indicate that
%% the nearby authors are not part of surrounding collaborations.

%% Mark off the abstract in the ``abstract'' environment. 
\begin{abstract}
The sun occasionally undergoes the so-called grand minima, in which its magnetic activity, measured by the number of sunspots, is suppressed for decades. The most prominent grand minima, since the beginning of telescopic observations of sunspots, is the Maunder minimum (1645--1715), when the sunspots became rather scarce. The mechanism underlying the grand minima remains poorly understood as there is little observational information of the solar magnetic field at that time. In this study, we examine the records of one candidate aurora display in China and Japan during the Maunder minimum. The presence of auroras in such mid magnetic latitudes indicates the occurrence of great geomagnetic storms that are usually produced by strong solar flares. However, the records of contemporary sunspot observations from Europe suggest that, at least for the likely aurora event, there was no large sunspot that could produce a strong flare. Through simple theoretical arguments, we show that this geomagnetic storm could have been generated by an eruption giant quiescent filament, or a series of such events. 

\end{abstract}

%% Keywords should appear after the \end{abstract} command. 
%% See the online documentation for the full list of available subject
%% keywords and the rules for their use.

%% From the front matter, we move on to the body of the paper.
%% Sections are demarcated by \section and \subsection, respectively.
%% Observe the use of the LaTeX \label
%% command after the \subsection to give a symbolic KEY to the
%% subsection for cross-referencing in a \ref command.
%% You can use LaTeX's \ref and \label commands to keep track of
%% cross-references to sections, equations, tables, and figures.
%% That way, if you change the order of any elements, LaTeX will
%% automatically renumber them.
%%
%% We recommend that authors also use the natbib \citep
%% and \citet commands to identify citations.  The citations are
%% tied to the reference list via symbolic KEYs. The KEY corresponds
%% to the KEY in the \bibitem in the reference list below. 

\section{Introduction} \label{sec:intro}

Solar magnetic activity varies with various timescales \citep{Hathaway2010, Usoskin2017}. 
Among these, are periods called grand minima, in which magnetic activity is 
significantly suppressed. The Maunder minimum (1645-1715) is the most prominent grand minimum 
since the beginning of the telescopic observations of the sun \citep{Eddy1976}. 
The dynamo mechanism of the grand minima is still controversial \citep{Charbonneau2010}, 
but the states of the solar magnetic field and the heliosphere during the Maunder minimum have 
been inferred by multiple proxies \citep{Kataoka2012, Usoskin2015, Riley2015}.
Analyses of cosmogenic radionuclides and sunspot records suggest that the Schwabe cycle 
persisted even during the Maunder minimum 
\citep{Ribes1993, Beer1998, Miyahara2004, Vaquero2015}. 
Another useful proxy for past solar activity is the record of the aurora borealis. 
Quite a few catalogues of the past aurora records were compiled by various authors, 
as summarized in Chap. 6.5 of \citet{Vaquero2009}. 

It should be noted that three proxies of past solar activity, namely, 
sunspots, cosmogenic isotopes, and midlatitude auroras, reflect 
various aspects of the solar magnetic activities. 
The sunspot number seems to be the most direct indication of 
a solar magnetic field, but sunspots appear 
only when a sufficiently large magnetic flux ($\gtrsim 10^{19}$ Mx) is concentrated 
enough to suppress convective energy transport \citep{Leka1998}. 
A large part of the solar surface is covered with sparse but significant 
magnetic flux, which does not appear as a sunspot \citep{deWijn2009}. 

Cosmogenic radionuclides are created by the precipitation of high-energy particles from space. 
Their correlation with the solar magnetic field is rather complex \citep{Usoskin2017}. 
When the sun is magnetically active, the heliosphere is filled with 
a strong magnetic field that suppresses the intrusion of galactic cosmic rays into the heliosphere. 
Thus, the production of cosmogenic isotopes in the terrestrial atmosphere is low. 
The production rate of radionuclides is also modulated by the secular variation of the geomagnetic field. 
Moreover, energetic particles produced by intense solar flares/eruptions and 
resultant interplanetary shocks also precipitate into the Earth's atmosphere and 
produce radionuclides \citep{Usoskin2006, Usoskin2012, Barnard2018}. 

The aurora borealis in the mid- to low-latitudinal regions can be regarded 
as a proxy for large eruptive events in the sun. 
The physical connections between the solar eruptions and the low-latitude auroras are 
rather complex, but the widely accepted outline is as follows: 
Eruptive events in the sun such as flares and filament eruptions launch 
coronal mass ejections (CMEs) into interplanetary space. 
When the southward component of the magnetic field is embedded in shock sheaths and/or the interplanetary CME (ICME),
its energy is effectively transferred to the magnetosphere via magnetic reconnection, 
which enhances the magnetospheric convection, resulting in the development of 
the ring current and the equatorward expansion of the auroral oval \citep{Akasofu1963}.
Empirically, it is known that the latitudinal extent of the aurora oval is correlated with the Dst index, 
which is a measure of the ring current development \citep{Yokoyama1998}, 
and the peak Dst index is correlated with the strength of 
the southward component of the magnetic field and the velocity of the solar wind. 
Thus, the auroral displays in mid- to low-latitude magnetic latitude can be regarded 
as evidence of strong solar eruptions. 
However, it should be noted that a strong solar eruption does not cause 
mid- to low-latitude aurora when the eruption is not toward the Earth, 
as in the near-miss extreme ICME in 2012 \citep{Baker2013}, 
or its magnetic field is not southward, 
as in the extremely fast ICME episode on 1972 August 4 \citep{tsurutani2013}. 

It is often difficult to confirm whether an aurora-like description 
in a historical document actually refers to an aurora display or 
a different phenomenon such as atmospheric optics \citep[e.g.,][]{UsoskinEA2017, Neuhaeuser2018}. 
Supplementary information, such as time, directions, and the moon phase, 
is useful in examining the probability that a description is actually 
a record of an aurora display. Namely, a record is certainly not a record of an aurora display
if the display was observed during daytime. Moreover, a record is more likely to be a record of an aurora display
if the display was observed in the northern sky (in the case of the northern hemisphere) 
and the moon phase is close to the new moon phase. 
\citep{Kawamura2016}.
\citet{Neuhaeuser2015} proposed five criteria for the likeliness of a record to be that of an aurora: color, aurora-typical motion, direction (northward), night-time observations and repetition of the event. 
While these criteria are apparently useful to evaluate the likeliness of the aurora records, 
it should be noted that they are not the necessary conditions to be an auroral record. 
For instance, a bright aurora can be seen even in the presence of a bright moon 
during extremely intense geomagnetic storms such as 
the great auroral event in 1847 \citep{Cliver2013}. 
Likewise, auroral display may be seen equatorward during such intensive magnetic storms.  
If the equatorial boundary of the aurora oval extended down to a lower as low as $\sim 30$ degree magnetic latitude (MLAT) 
\citep{Green2006, Hayakawa2016, Ebihara2017, Hayakawa2017b, Hayakawa2018a, Hayakawa2018b, Hayakawa2019a, Love2018}, 
the aurora was visible in the equatorward sky from midlatitude (say, $\sim$ 50 degree) regions. 
\citet{Stephenson2019} compared the criteria proposed by \citet{Neuhaeuser2015} with observational evidence and doubted their validity with counter-examples.
Simultaneous and independent observations at distant locations 
provide reasonable support for the interpretation of a display as an aurora
because it can exclude the possibility of local phenomena, such as 
atmospheric scatterings, and fiction by the author \citep{Willis2000}. 

The aurora records during the Maunder minimum, 
particularly those from northern to central Europe, 
have been used to infer the degree of solar magnetic activity during the period \citep{Eddy1976, Usoskin2015, Riley2015}. 
There are also Chinese, Korean, and Japanese records 
that can be regarded as candidates for auroral displays 
in these regions \citep{Yau1995, Xu2000, Lee2004, Kawamura2016}, 
where the geographic latitude ($30 \sim 40$ degree) 
is lower than that of central Europe and the magnetic 
latitude is even lower ($20 \sim 30$ degree). 
However, it is uncertain how much of them are the record of true aurora 
\citep{UsoskinEA2017, Neuhaeuser2018}. 

To the best of our knowledge, the only set of simultaneous, 
and hence very probable, records of auroral display 
in East Asia during the Maunder minimum 
are from March 1653, as pointed out by \citet{Willis2000}. 
There is another candidate from 1672, but as discussed in 
the next section, it was a result of dating error.
While there still exist other records that are potentially of aurora display in the mid-latitudes, 
so far this event is the most likely one during the Maunder minimum.
Naturally, the question arises as to how such a strong geomagnetic 
storm was produced from the sun in its grand minimum. 
In this study, we address this question.

In section \ref{sec:record}, we examine East Asian 
aurora records and contemporary sunspot records from Europe, 
and verify that when the March 1653 aurora was seen, the sun was 
almost spotless. 
In section \ref{sec:theory}, we show that such a low-latitude aurora 
can be produced by an eruption of large quiescent filaments,
by examining some empirical models. 
Implications of the study will be discussed in section \ref{sec:summary}.

\section{Sunspot and aurora records}
\label{sec:record}
\subsection{Simultaneous aurora records in East Asia}

\citet{Willis2000} reported two simultaneous observations of auroral displays 
on March 2, 1653: one from China and one from Japan. 
The moon phase of the day was 0.56. 
We revisited the original records of these historical reports and provide their translation, observational site, and references as follows.

\begin{enumerate}
\item 
On 1653 March 2, during night, fiery lights illuminated the heaven in four directions. After a while they became a blue wisp. [Y\v{a}nzh\={o}uf\v{u} C\'{a}oxi\`{a}nzh\`{i}, v.18, f.10a]

\item 
On 1653 March 2, ... recently, red and white vapours appeared between Nasu and Odawara in Shimotsuke. They were like flags. It is said that the red vapour vanished earlier. [Tokugawa Jikki, v.11, p.70]

\end{enumerate}

The geographic location of the first record, C\'{a}oxi\`{a}n, is 
 N$34^{\circ}$59', E115$^{\circ}$32'. 
 Using the magnetic field model GUFM1 \citep{Jackson2000}, 
 the MLAT is calculated to be N28.4. 
The location of the observation of the second record from Japan is described as 
``between Shimotsuma, Nasu, and Odawara.'' 
Nasu is located at N37$^{\circ}$01', E140$^{\circ}$07', corresponding to
MLAT = N29.9. Odawara is located at N36$^{\circ}$52', E140$^{\circ}$01', 
corresponding to MLAT = N29.7.

In addition to the March 1653 event, an additional candidate, 
from a set of simultaneous auroral observations in East Asia 
during the Maunder minimum, can be found in the literature. 
This candidate is from records of September 1672. 
One is a record in Ji\={a}ngs\={u} P\'{i}ngw\`{a}ngzh\`{i} (v.13, f.6b): 
``On 1672 September 21, during night, crimson vapor filled the heaven like fiery sand.'' 
Another candidate can be found in the catalogue of astronomical records in Japan,
compiled by \citet{Osaki1994}, and also in \citet{Nakazawa2004}, who presumably 
referred to Osaki's catalogue. According to these publications, there is 
a record of ``Heaven was red'' ({\it TEN-AKA-shi} in Table 1 of \citet{Nakazawa2004} )" on September 17, 1672. 
However, the contemporary woodprint 
{\it Honcho-Nendai-ki (v.5, f.42b)} dated this record as September 7, 1935. 
It seems that \citet{Osaki1994} misread the year, as he recorded 
the year as the 12th year of the {\it Kanbun} period, corresponding to 1672, 
which was actually the 12th year of the {\it Kan'ei} period (1635) in the original woodprint.

\subsection{Sunspot records}

While two independent records, dated March 2, 1653, at distant locations 
are suggestive of an aurora display 
in such low-latitude regions, we need more independent records to decisively prove this. 
Furthermore, if these are indeed real aurora records, the question arises as to whether there were large sunspots that could produce such geoeffective eruptions. 
To examine the level of solar activity at the time of the March 2, 1653 event, 
we consult the contemporary sunspot record by \citet{Hevelius1673}, examined by \citet{Carrasco2015}. 
In 1653, there were records of one day in February and seven days in March:

\begin{enumerate}
\item Feb. 5 {\it Macula exigua apparuit} [Small spot appeared]
\item Mar. 1 {\it Nil Macularum} [No spots]
\item Mar. 9 {\it Nil Macularum} [No spots]
\item Mar. 23 {\it Macula in quadrante Orientali Solis} [Sunspot in the eastern quadrant of the Sun]
\item Mar. 24 {\it Macula Solis} [Sunspot]
\item Mar. 25 {\it Binae maculae in Sole} [Double spots in the Sun]
\item Mar. 27 {\it Iam penitus disparuerant} [They had completely disappeared]
\item Mar. 29 {\it Nil Macularum} [No spots]
\end{enumerate}

The solar eruption that produced the March 2 aurora should have occurred a day or a few days before. 
According to Hevelius' record, there was no sunspot on March 1. 
Does this indicate that the solar eruption that caused the March 2 aurora occurred in the spotless sun?
It is possible that there were sunspots at the end of February, which produced the geoeffective eruption, 
but they disappeared before Hevelius' observation on March 1. 
If the sunspots disappeared from the solar surface within a few days, 
they must have been very small or almost completely decayed when the eruption occurred. 

Another possibility is that the sunspots were very close to the sun’s west limb when 
they produced the eruption, and they rotated behind the limb before March 1.
Most of the extreme geomagnetic storms are produced by eruptions from large active regions 
\citep[though there are exceptions such as the great aurora on September 25, 1909, see][]{Hayakawa2019a} 
that are likely to remain visible one rotation later. 
The records of sunspots from March 23--25 may correspond to this idea, 
but if the sunspots were close to the west limb at the end of February and remained present 
one rotation later, they should have appeared in the western hemisphere on March 23. 
This is not consistent with Hevelius' record. 

Thus, Hevelius' record suggests that, when the eruption that caused the March 2, 1653 aurora occurred,
there were no sunspots present on the solar surface, or only small sunspots existed. 
However, we note that it is often not straightforward to interpret the historical sunspot records \citep{Munoz-Jaramillo2018}, and our argument is of speculative nature.

\section{Physical interpretations}
\label{sec:theory}

\subsection{Intensity of geomagnetic storms}
It is known that the equatorward extension of the aurora belt is correlated with 
the Dst index, a standard measure of the ring current energy. 
\citet[hereafter Y98]{Yokoyama1998} examined 423 geomagnetic storms and 
found that the corrected geomagnetic latitude $\Lambda$ of the equatorward boundary of the auroral oval decreases with the peak Dst index. Y98 assumed that the magnetic field line is the dipole, and that $\Lambda$ is associated with McIlwain's L-value \citep{McIlwain1961} as
\begin{eqnarray}
L = \frac{1}{\cos^2\Lambda}.
\end{eqnarray}
Hereinafter, we refer $\Lambda$ to as the magnetic latitude unless otherwise mentioned.
In the following we use the scaling of Y98 to estimate the Dst index of the March 2, 1653 event.

For this purpose, we need to convert the location of the observation site 
into the magnetic latitude of the equatorward boundary of the aurora. 
From simple geometry, one can derive the following equation: 
\begin{equation}
(R+h) \cos (\Lambda - \Lambda_0) = R + (R+h) \sin (\Lambda - \Lambda_0) \tan \theta, 
\end{equation}
where $R$ is the Earth's radius, $h$ is the altitude of the upper boundary of the aurora, 
$\Lambda_0$ is the magnetic latitude of the observational site of the aurora, 
and $\theta$ is the elevation angle of the aurora. 
Unfortunately, there is no information on the elevation angle (such as constellations) 
for this event, so we assume a conservative value of $\theta \sim 10^{\circ}$. 
Substituting $\Lambda_0 \sim 29^{\circ}$, a typical altitude of red aurora $h \sim 500$ km,
and $R \sim 6400$ km, we obtain $\Lambda \sim 43^{\circ}$.

In terms of Dst and $\Lambda$, the scaling of Y98 can be defined as 
\begin{equation}
{\rm Dst} \approx 12 - 2200 \cos^6 \Lambda \, {\rm (nT)}
\end{equation}
This value of $\Lambda$ corresponds to Dst $\sim$ -325 nT, which can be regarded as a conservative value because this equation was derived on the basis of 423 geomagnetic storms. Y98 also suggested the other equation as
\begin{equation}
  {\rm Dst} \approx 60 -3400 \cos^6 \Lambda \, {\rm (nT)},
\end{equation}
which was derived on the basis of the main phase of the extremely large magnetic storm of March 1989. With this equation, Dst is estimated to be $\sim$ -460 nT, which may be regarded as the lower limit of Dst. Hereinafter, we take the conservative value of Dst, that is, -325 nT.

\subsection{Possible solar origins}

What are the solar origins that are capable of driving Dst $<$ -300 nT storms in the absence of a large active region? 
Fast solar winds in corotating interaction regions (CIRs) can drive 
geomagnetic storms, but they usually do not exceed Dst = -120 nT, and 
their maximum strength is likely to be $\sim$ -160 nT \citep{Richardson2006}. 

Another possibility is CMEs from eruptions of quiescent filaments. 
Strong geomagnetic storms without any association with major flares were recognized 
as ``problem storms'' \citep{Dodson1964}. Later, they were found to be associated with eruptions 
from the quiet sun, often accompanied with the disappearance (eruption) of a dark filament
 \citep{Joselyn1981, McAllister1996, Cliver2009, Zhang2007}. 

\citet{McAllister1996} studied in detail one such geomagnetic storm event that occurred in April 1994. 
The event was associated with a giant soft X-ray arcade formation in the quiet sun, on April 14, 1994,
which eventually produced an intense geomagnetic storm (peak Dst of -201 nT) and great aurora on April 17. 
Although the filament was barely visible, the giant arcade seen in the soft X-ray 
was considered to be the aftermath of an eruption from a large-scale magnetic neutral line 
surrounding the polar region of the sun \citep{Tsuneta1992, Tripathi2004}. 

\begin{figure}
\plotone{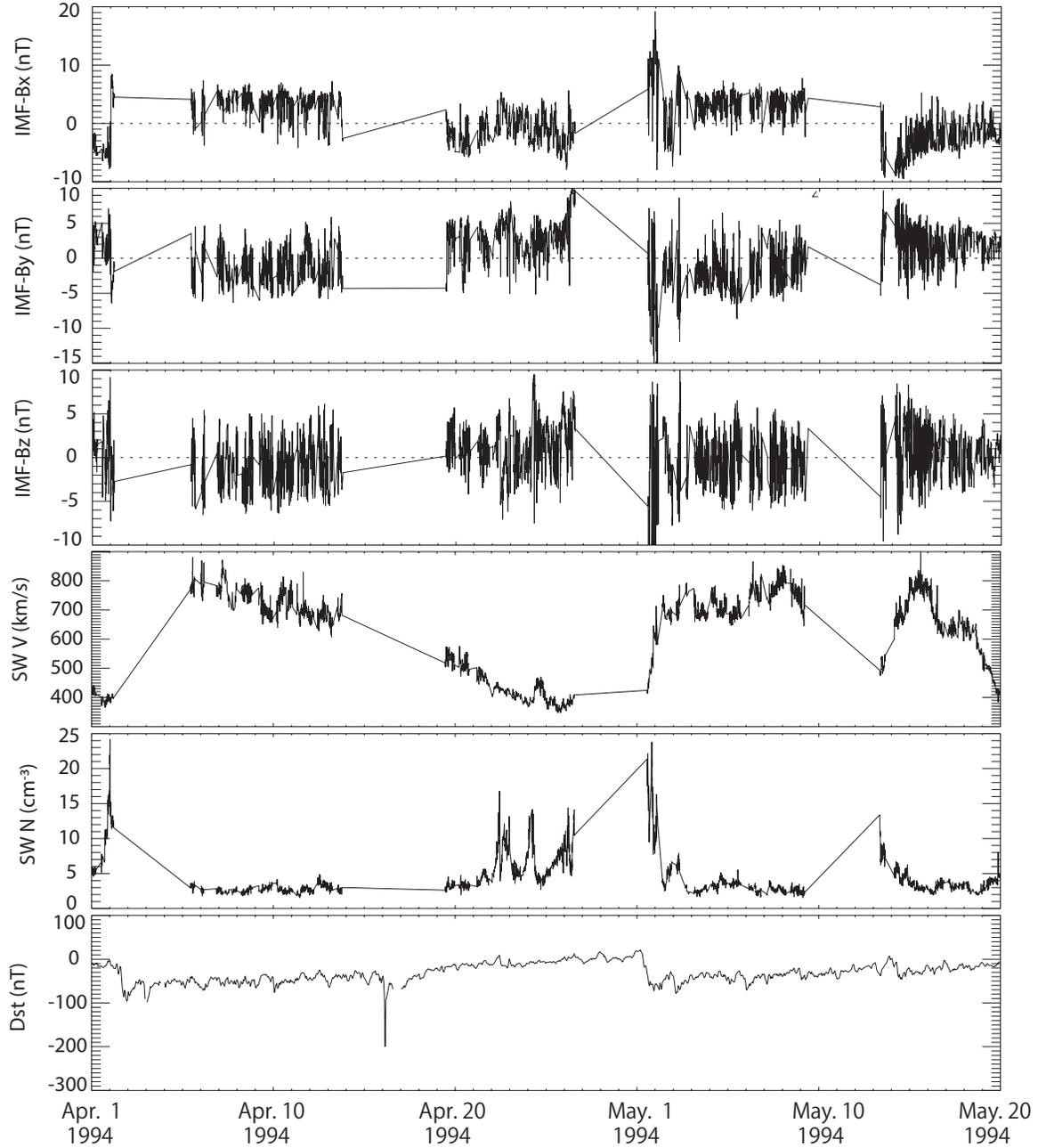}
\caption{Three components of magnetic field in the geocentric solar magnetic coordinate, velocity and density in the solar wind measured by IMP8 and the Dst index in April 1994. \label{fig:1994Apr}}
\end{figure}

Figure \ref{fig:1994Apr} shows the solar wind parameters (three components of magnetic field, velocity and density) obtained by the IMP8 satellite and the Dst index 
in April 1994. Unfortunately, there was no record of data during the storm, but one can see that the solar wind 
before the onset of the storm was relatively fast ($\sim$ 700 km s$^{-1}$). 
Although one cannot see the ICME in this figure, it was observed by Ulysses at 3AU \citep{McAllister1996}. 
The average speed of the CME, calculated from the onset times of the X-ray arcade 
and the geomagnetic storm, is about 600 km s$^{-1}$.
An interesting feature, as shown in Figure \ref{fig:1994Apr}, is that the Dst index was already about -50 nT 
before the start of the geomagnetic storm, suggesting a 
 long-lasting moderate activity in the magnetosphere due to the Russell-McPherron effect \citep{Russell1973}. 
 This is evident by the long-lasting toward magnetic field ($Bx >> 0$) and the date of this event being close to the spring equinox. The high solar wind speed ($\sim$700-800 km s$^{-1}$) might participate in the long-lasting moderate activity in the solar wind.

During the November 1991 event reported by \citet{Cliver2009}, the Dst index of the geomagnetic 
storm was as low as -354 nT. This is the strongest known geomagnetic storm associated with 
the eruption of a quiescent filament. \citet{Cliver2009} noted that there was a nearby active region, 
and hence, the filament may not have arisen from a ``pure'' quiet sun. This is possibly related 
to the relatively high speed ($\sim$ 1000 km s$^{-1}$) of the CME whose origin was a quiescent filament eruption. 

In general, CMEs from a quiet sun are slower. 
According to a statistical study by \citet{Yashiro2009}, the upper limit of the CME speed is well correlated 
with the peak flux and fluence of soft X-rays, and the speed of the CME associated with the GOES B-class 
(peak flux lower than 10$^-5$ W m$^{-2}$) is mostly below 1000 km s$^{-1}$ (see Figure 6 of \citet{Yashiro2009}). 
On the other hand, the field strength of the magnetic cloud of such slow CMEs can be as large as 30--40 nT \citep{Owens2005}. 

Putting these empirical results together, we hypothesize that the March 2, 1653 event was 
driven by an eruption (or a series of multiple eruptions) from a quiet sun, which 
produced an interplanetary CME with a relatively slow speed ($V \sim 600$ km s$^{-1}$) and a
strong southward magnetic field ($B \sim 30-40$ nT).

\subsection{Validity of solar wind parameters}

To quantitatively examine whether such an eruption, as discussed in the previous subsection, can drive a geomagnetic storm 
as intense as Dst $<$ -300 nT, we consider the empirical evolutionary equation of the Dst index 
derived by \citet{Burton1975}.
\begin{equation}
\frac{d Dst*}{dt} = Q(t) - \frac{Dst*}{\tau_{decay}}.
\label{eq:Burton}
\end{equation}
Here, $Q(t) \propto V_{SW}B_s$, where $V_{SW}$ is the solar wind velocity and $B_s$ is the southward component 
of the interplanetary magnetic field, $Q$ is the energy injection rate, and $\tau_{decay}$ is the decay time.

Note that $Dst*$ is a corrected version of $Dst$ \citep{Burton1975, Obrien2000a} given by
\begin{equation}
Dst* = Dst -b \sqrt{P_{dyn}} + c, 
\end{equation}
where $P_{dyn}$ is the solar wind dynamic pressure, and $b$ and $c$ are constants. 
Because the second and third terms on the right-hand side are of the order of 10 nT, 
we neglect these terms as they do not affect the analyses of this study. 
Therefore, we omit the asterisk in the following. 

Various models of energy injection rate $Q(t)$ and decay time $\tau_{decay}$ have been proposed 
\citep[See review in][]{Obrien2000a}. 
\citet{Obrien2000b} compared different models of $Q(t)$ and $\tau_{decay}$ 
and found that the following model (AK2 model) showed the best performance in reproducing 
the observations. 
\begin{equation}
Q(t) = -4.4 (V_{SW}B_s -0.5)
\end{equation}
and 
\begin{equation}
\tau_{decay} = 2.4 \exp \Big( \frac{9.74}{4.69 +V_{SW}B_s } \Big).
\label{eq:tdecay}
\end{equation}
Here, the electric field $V_{SW}B_s$ is in mV m$^{-1}$, and $\tau_{decay}$ is in hours. 
Namely, the decay time is shorter for larger energy injections from the solar wind. 

When $V_{SW}B_s$ is constant in time, the evolution of $Dst$ stops when 
the injection and decay terms balance. In an actual situation, $V_{SW}B_s$ varies with time. 
For simplicity, we assume that $Q(t)$ is constant during $0 \le t \le \tau_{dur}$ and set to zero 
when $\tau_{dur} < t$. 

Because we consider only slow CMEs by quiescent filament eruptions, 
we fix $V_{SW}$ to 600 km s$^{-1}$ and change $B_s$ and $\tau_{dur}$. 

\begin{figure}
\plotone{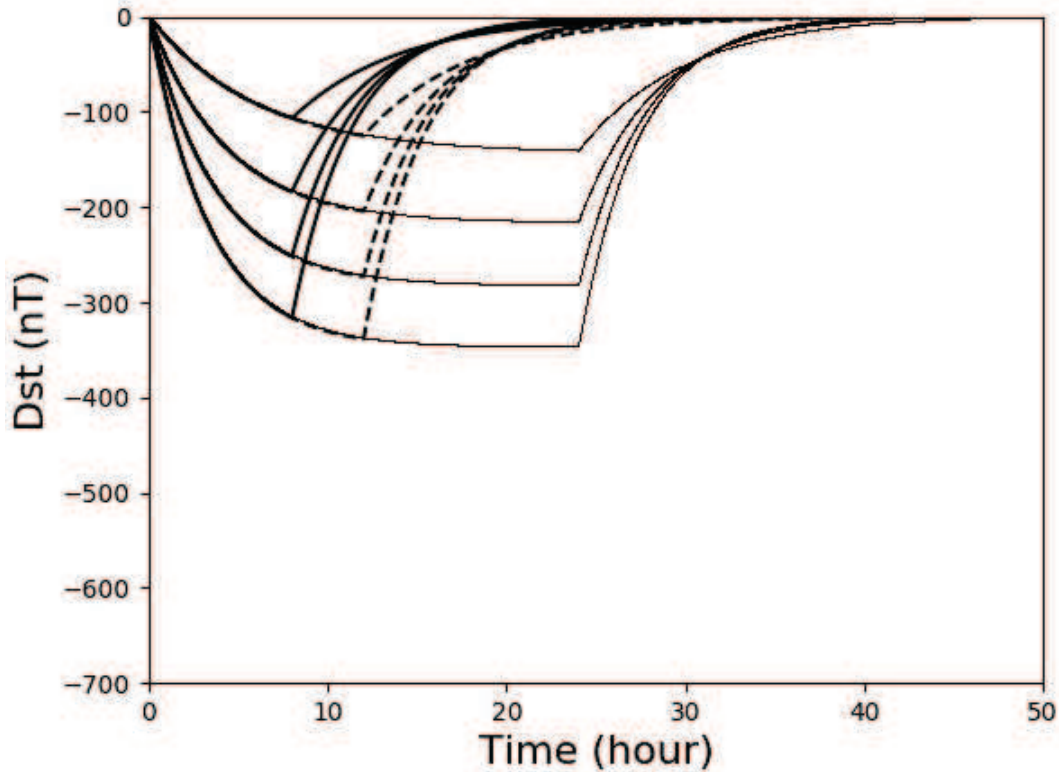}
\caption{Evolution of Dst index for $V_{SW}$ = 600 km s$^{-1}$; $B_s = 10, 20, 30, 40$ nT, $\tau_{dur} = 8, 12, 24$(h), and AK2 model for $\tau_{decay}$. Thick, dashed, and thin lines correspond to the solutions for $\tau_{dur} = 8, 12, 24$(h), respectively,  and the lines with faster decrease in Dst correspond to the solutions for larger $B_s$. 
 \label{fig:Dst-AK2}}
\end{figure}

\begin{figure}
\plotone{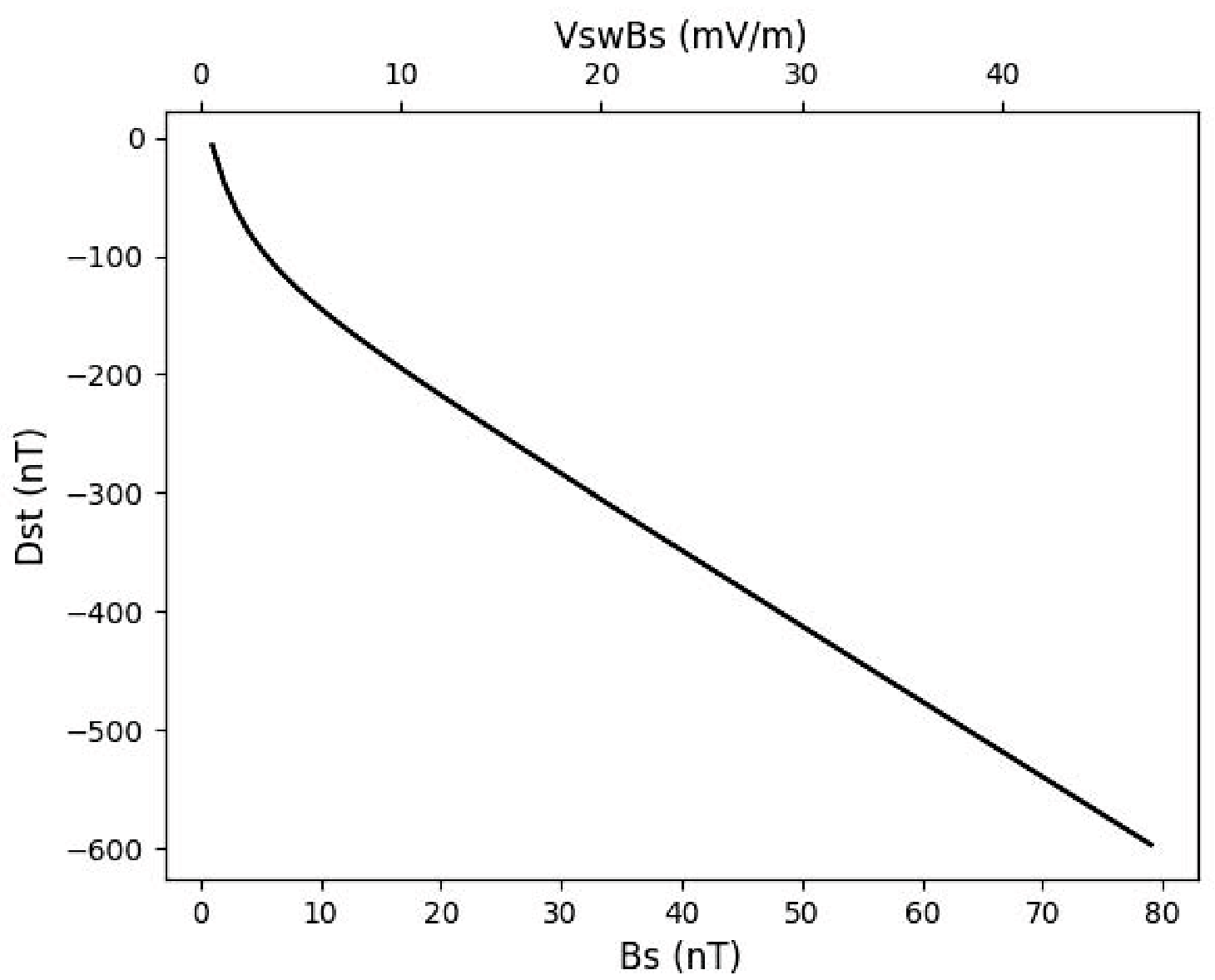}
\caption{Saturation value of Dst as function of southward component of solar wind magnetic field $B_s$ 
or associated electric field $V_{SW}B_s$ for AK2 model. $V_{SW}$ = 600 km s$^{-1}$ is assumed. 
 \label{fig:PeakDst}}
\end{figure}

Figure \ref{fig:Dst-AK2} shows the numerical solutions of equation (\ref{eq:Burton}) 
for $B_s = $ 10, 20, 30, and 40 (nT) and $\tau_{dur}$ is 8, 12, and 24 (h). 
$t_{decay}$ calculated by equation (\ref{eq:tdecay}) is 6.0, 4.3, 3.7, and 3.4 (h) 
for $B_s = $10, 20, 30, and 40 (nT), respectively. 
The dashed, solid, and dotted lines are the solutions when $\tau_{dur}$ is 8, 12, and 24 (h), respectively. 
The solutions with a faster decrease in the Dst correspond to a larger value of $B_s$. 
An obvious feature seen in Figure \ref{fig:Dst-AK2} is that the Dst almost saturates at around 
$t \sim 10$ (h) even if the energy injection continues. 
The saturation is a result of the balance between the energy injection term $Q(t)$ 
and the decay term $Dst/\tau_{decay}$ in equation (\ref{eq:Burton}), 
thus the saturation value $Dst_{sat}$ is given by
\begin{equation}
Dst_{sat} = Q(t) \tau_{decay} = -10.6 (V_{SW}B_s -0.5) \rm{exp}\big( \frac{9.74}{4.69 + V_{SW}B_s}\big)
\end{equation}
and is plotted in Figure \ref{fig:PeakDst}.
From Figures \ref{fig:Dst-AK2} and \ref{fig:PeakDst}, we conclude that, 
according to the AK2 model of \citet{Obrien2000a}, the Dst can exceed -300 nT 
if $B_s \gtrsim 30$ nT and $t_{dur} \gtrsim 8$ h.

It should be noted that the AK2 model in \citet{Obrien2000b} is based on the statistics of 
moderate (Dst $> -150$ nT) geomagnetic storms, although it reasonably 
reproduced the Dst evolution of an intense (Dst $<-200$ nT) storm \citep{Obrien2000a}. 
Whether the AK2 model (in particular, the form of $\tau_{decay}$) is relevant for 
extremely intense storms, such as the March 2, 1653 event, is not fully understood, 
primarily because of the scarcity of such data. 
Therefore, we also examine the effect of different $t_{decay}$. 
As a reference, in the absence of the decay term the solution of equation (\ref{eq:Burton}) is a simple linear function of $t$:  
\begin{equation}
{\rm Dst}(t) \approx -830 \Big( \frac{B_s}{40 \, \rm{nT}} \Big) \Big( \frac{V_{SW}}{ 600 \, \rm{km \, s}^{-1} } \Big) \Big( \frac{t}{ 8 \, \rm{h}} \Big) 
\end{equation}
for $t \leq \tau_{dur}$.

\begin{figure}
\plotone{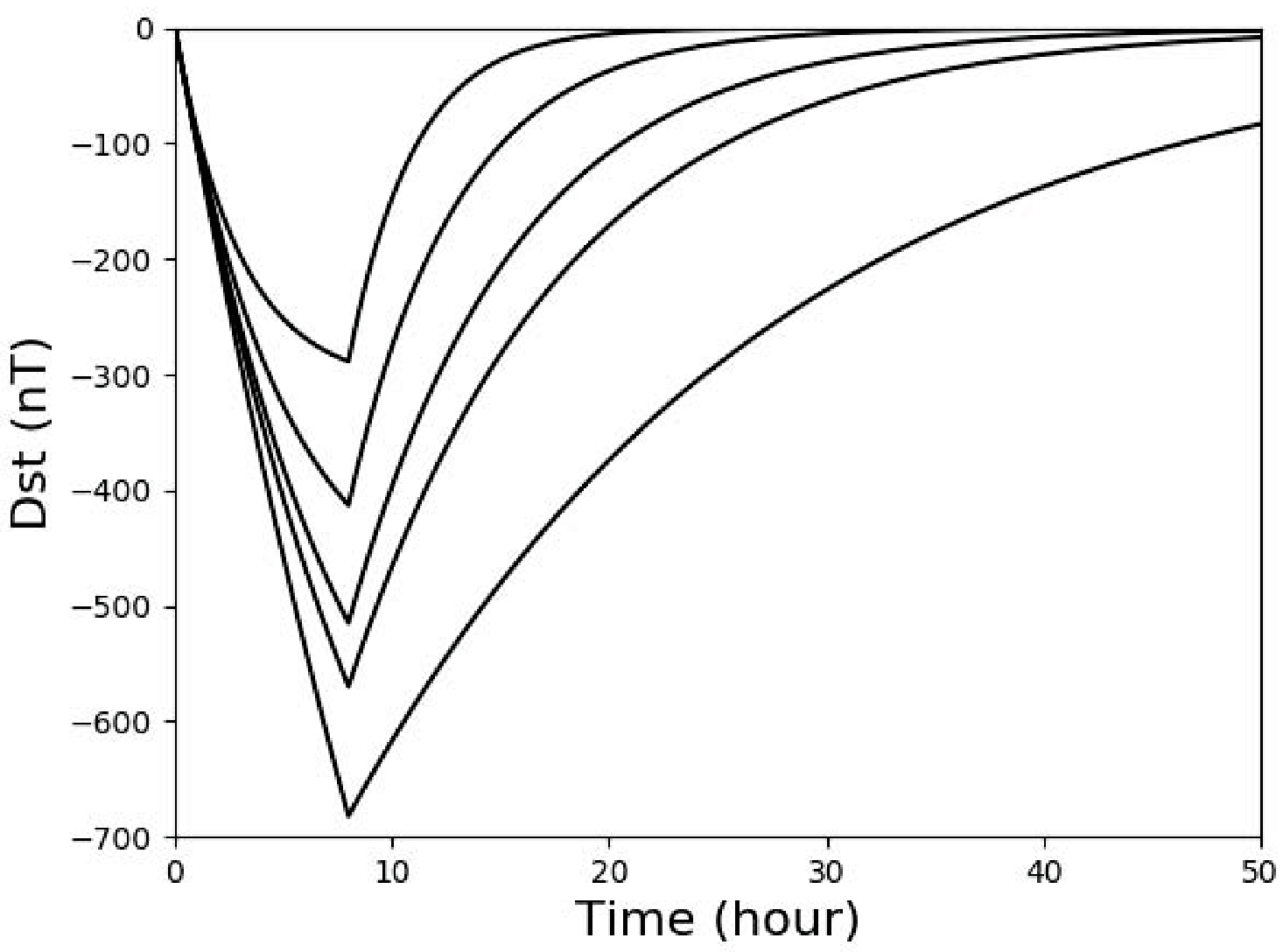}
\caption{Evolution of Dst index for $B_s = 40$ nT, $\tau_{dur} = 8 $(h), and 
$\tau_{decay} =$ 3, 5, 7.7, 10, and 20 (h). 
 \label{fig:Dst-tdec}}
\end{figure}

Figure \ref{fig:Dst-tdec} shows the numerical solution of equation (\ref{eq:Burton}) for $B_s = 40$ nT, $\tau_{dur} = 8 $(h), 
and $\tau_{decay} =$ 3, 5, 7.7, 10, and 20 (h). One can see that if $\tau_{decay}$ is sufficiently long, then
the peak Dst can reach as low as -700 nT.

Finally, we examine the validity of our hypothesis in terms of the magnetic flux. 
The total magnetic flux content in an ICME can be estimated from
in situ observations by assuming models of the magnetic structure of the ICME \citep[e.g., ][]{Dasso2007}. 
For this study, however, an order-of-magnitude estimate is sufficient. 
Using the characteristic length $V_{SW}\tau_{dur}$ and the magnetic field strength $B_s$, 
the magnetic flux $\Phi$ associated with the CME is roughly given by 
\begin{equation}
\Phi \approx B_s V_{SW}^2 \tau_{dur}^2 \approx 1.2 \times 10^{21} \Big( \frac{B_s}{ 40 \, \rm{nT}} \Big) \Big( \frac{V_{SW}}{ 600 \, \rm{km \, s}^{-1} } \Big)^2 \Big( \frac{\tau_{dur}}{ 8 \, \rm{h}} \Big)^2 \, {\rm Mx.}
\end{equation}
For an average magnetic field strength of 1 G, which is typical in the quiet-sun photosphere, 
$\Phi = 1.2 \times 10^{21}$ Mx corresponds to an area of $1.2 \times 10^{21}$ cm$^2 \sim$ 4\% of the solar hemisphere. 
It is a significant fraction of the solar surface but not an irrelevant value for the magnetic flux 
associated with a large-scale eruption from a quiet sun, such as that in April 1994.

\section{Discussion}
 \label{sec:summary},

In addition to the sunspot records and cosmogenic radioisotopes, 
the historical records of aurora displays during the Maunder minimum 
provide unique and independent information on the solar magnetic activity during this period. 
To the best of our knowledge, the simultaneous records in China and Japan on March 2, 1653 are the only 
known information of aurora displays during the Maunder minimum securely confirmed by simultaneous observations. 
The empirical correlation between the latitude of the aurora and the Dst index indicates that 
the geomagnetic storm of the March 2, 1653 event was extremely intense (Dst $\lesssim$ -300 nT).
From contemporary sunspot records and through some simple theoretical arguments, 
we propose that the sun was either spotless or with only tiny spots, and that the geomagnetic storm was driven by the eruption of quiescent filament(s).

If the proposed hypothesis is true, its significance is twofold. 
From the viewpoint of space weather, it indicates that even when the sun looks very quiet, 
serious space-weather hazards may occur, although their probability is low.
From the viewpoint of solar magnetism, it provides additional constraints on 
the magnetic configurations in the sun during the grand minimum. 

Here, we discuss additional factors that may enhance the geoeffectiveness of the eruption.
The first possible factor is multiple eruptions and CMEs. 
It has been known that consecutive eruptions can be more geoeffective than 
an isolated CME \citep{Gopal2004, Kataoka2017}. 
Such consecutive CMEs are usually produced by flare-productive active regions 
\citep{Chatterjee2013}. 
To our knowledge, there are no known intense geomagnetic storms driven by 
multiple and consecutive eruptions purely from a quiet sun. 
However, consecutive eruptions that involve quiescent filaments have also been observed 
and are known as ``sympathetic eruptions'' \citep{Schrijver2011, Titov2012}. 

Another solar factor that may enhance the geoeffectiveness is a fast solar wind from a disk coronal hole. 
\citet{Asai2009} investigated the active region, NOAA 10798, that emerged in the middle of a disk coronal hole and produced 
three moderate (M-class) flares. The associated CMEs were very fast ($1200 \sim 2400$ km s$^{-1}$) 
because they were in the fast solar wind from the coronal hole. In addition, the faster CME caught up to the preceding, 
relatively slow CME, and eventually they merged.
The combination of fast velocity and the fact that the two consecutive CMEs merged before they arrive at the Earth,
enhanced the geoeffectiveness, and consequently an intense geomagnetic storm (Dst = -216 nT) occurred. 
While the assistance of a fast solar wind in the acceleration of CMEs appears to be a feasible mechanism, 
directly applying the same scenario to quiescent filament eruptions is problematic because 
it is unlikely that large-scale quiescent filaments exist in a large coronal hole,
as a large-scale filament requires a long magnetic neutral line while a large coronal hole requires a large unipolar region. 

The last factor we consider is the seasonal effect. 
It has been known that geomagnetic activity is stronger around the equinoxes than around the solstices, 
and indeed the March 2, 1653 event occurred not far from the spring equinox. 
The proposed mechanisms that account for these seasonal variations are 
the axial hypothesis \citep{Cortie1912, Bohlin1977}, 
equinox hypothesis \citep{Bartels1932, Svalgaard1977}, 
and Russell-McPherron effect \citep{Russell1973}. 
The seasonal variation is usually measured by the occurrence rate of the geomagnetic storms or by indices such as {\it aa} and Dst indices, and not necessarily by the intensity of the individual storm. 
However, if the magnetosphere was already disturbed to some extent, it may help 
to enhance the geomagnetic storm driven by CMEs. 
The axial hypothesis considers that the Earth is in the heliograhic latitude near the equinoxes,
where the solar wind condition can be more hazardous owing to midlatitude sunspots \citet{Cliver2000} 
or coronal holes \citep{Bohlin1977}. For the March 2, 1653, event, there was no sunspot 
but the effect of the fast solar wind at a higher heliographic latitude may have played a role. 
The equinox hypothesis attributes the seasonal variation to the varying angle 
between the Earth's magnetic dipole axis and the Sun-Earth line. 
According to \citet{Cliver2000}, this effect significantly contributes to the seasonal variation 
but primarily suppresses the coupling efficiency of the solar wind and the magnetosphere 
near solstices. Thus, it is not relevant as a mechanism for enhancing the geoeffectiveness 
of eruptions near equinoxes. 
Finally, in the Russell-McPherron effect, the solar wind magnetic field 
in a solar equatorial plane has a southward component in the geocentric 
solar magnetospheric coordinate near the equinoxes. 
This has been regarded as the principal cause of the enhancement of geomagnetic 
activity near the equinoxes \citet{Cliver2000, Zhao2012}. 
Although its contribution to the strongest storms is considered to be relatively minor \citep{Lookwood2016}, 
and it may have played some role in the May 2, 1653 event, too. 

We note that although the date was near the spring equinox, there is no direct evidence for the factors 
discussed above playing any role in the March 2, 1653 event. 
Future surveys of so-far-unknown records of aurora displays during the Maunder minimum 
is a promising way to bring new insights on the magnetic activity of the Sun in its grand minima.

\acknowledgments

This work was supported by Grant-in-Aids from the Ministry of Education, Culture, Sports, Science and Technology of Japan, Grant Number JP18H01254 (PI: H. Isobe), JP15H03732 (PI: Y. Ebihara) and a Grant-in-Aid for JSPS Research Fellow JP17J06954 (PI: H. Hayakawa). It was also supported by Kyoto University’s Supporting Program for the Interaction-based Initiative Team Studies “Integrated study on human in space” (PI: H. Isobe), the Mission Research Projects of the Research Institute for Sustainable Humanosphere (PI: H. Isobe) and SPIRITS 2017 (PI: Y. Kano) of Kyoto University.

%% This command is needed to show the entire author+affilation list when
%% the collaboration and author truncation commands are used.  It has to
%% go at the end of the manuscript.
%\allauthors

%% Include this line if you are using the \added, \replaced, \deleted
%% commands to see a summary list of all changes at the end of the article.
%\listofchanges

\end{document}